\documentclass[12pt]{iopart}

\usepackage{iopams}

\newcommand{\la}{\lambda}
\newcommand{\ga}{\gamma}
\newcommand{\vp}{\varphi}
\newcommand{\be}{\begin{equation}}
\newcommand{\ee}{\end{equation}}
\newcommand{\bea}{\begin{eqnarray}}
\newcommand{\eea}{\end{eqnarray}}
\newcommand{\nn}{\nonumber}

\begin{document}

\title[Characteristic polynomials in real Ginibre ensembles]{Characteristic
  polynomials in real Ginibre ensembles} 

\author{G. Akemann$^1$, M.J. Phillips$^1$, and H.-J. Sommers$^2$}

\address{\it $^1$Department of Mathematical Sciences 
\& BURSt Research Centre,\\
\ Brunel University West London,
UB8 3PH Uxbridge, United Kingdom\\
$^2$Fachbereich Physik, Universit\"{a}t Duisburg-Essen,
47048 Duisburg, Germany
} 
\ead{Gernot.Akemann@brunel.ac.uk, Michael.Phillips@brunel.ac.uk,
H.J.Sommers@uni-due.de}

\begin{abstract}
We calculate the average of two characteristic polynomials for the real Ginibre
ensemble of asymmetric random matrices, and its chiral counterpart. Considered
as quadratic forms they determine a skew-symmetric kernel from which all
complex eigenvalue correlations can be derived. Our results are obtained in a
very simple fashion without going to an eigenvalue  representation, and are
completely new in the chiral case. They hold for Gaussian ensembles which are
partly symmetric, with  kernels given in terms of Hermite and Laguerre
polynomials respectively, depending on an asymmetry parameter. This allows us
to interpolate between the maximally asymmetric real Ginibre and the Gaussian
Orthogonal Ensemble, as well as their chiral counterparts.

\end{abstract}

\pacs{
02.10.Yn,
0250.-r, 0540.-a
}

\section{Introduction}\label{intro}

Random Matrix Theory is known to enjoy a wide range of applications in the
physical sciences and beyond. This remains true when the 
eigenvalues of the operator to be described move into the complex plane.
However, the ensemble that is perhaps the most interesting of these, 
the real Ginibre ensemble \cite{Ginibre65} 
dealing with real-valued asymmetric matrix entries, 
has turned out to be the most difficult.
Possible applications of these ensembles include
neural networks \cite{Sompolinsky88}, directed Quantum Chaos 
\cite{Efetov97}, Quantum Chromodynamics \cite{HOV}, financial markets 
\cite{KWAPIEN06}, and quantum information theory \cite{Bruzda08}.

The mathematical difficulty in solving these ensembles is due to the fact that
they allow for combinations of both real and complex conjugate eigenvalue
pairs, with their characteristic equation having only real entries.
Apart from results on the spectral density 
\cite{Sommers88,EKS94,Efetov97} 
an eigenvalue representation \cite{Lehmann91,Edelman97} was derived as a
starting point for studying systematically 
higher order eigenvalue correlation functions. 
Only very recently was their Pfaffian structure explicitly revealed
\cite{Kanzieper05,Sinclair06}, and
the probability $p_{N,k}$ that an $N\times N$ matrix has exactly $k$ real
eigenvalues \cite{Kanzieper05,Akemann07} 
as well as all correlations for $k=0$ were computed 
\cite{Akemann07}. Finally the complete solution for all
real and complex eigenvalue correlations was achieved independently by three
different groups
\cite{Sommers2007,Forrester07,Borodin07,Forrester08,Sommers2008}. 

In this paper we will give a very simple derivation for the generating kernel 
of all complex eigenvalue correlations in the general so called elliptic case,
dealing with partly symmetric matrices depending on an asymmetry parameter.
We also present new results for the chiral real Ginibre ensemble as a
two-matrix model which has not yet been considered.

In the next section 
we will explain the relation between the complex
eigenvalue density and characteristic polynomials. After defining the elliptic
real Ginibre ensemble and its new chiral extension
we consider their corresponding results 
in two separate sections \ref{Gin1} and \ref{chGin1}.
Our conclusions are presented in section \ref{concs}.

\section{The r\^ole of characteristic polynomials generating the kernel}
\label{role}

We start with the simplest ensemble considered here, the real Ginibre ensemble
at maximal asymmetry. It is just given by a Gaussian measure in the space
of real, asymmetric  $N \times N$ matrices 
which is invariant under orthogonal transformations:
\begin{equation}
\label{dmuA} 
d\mu (A) \equiv \prod_{i,j=1}^{N} 
\left( \frac{dA_{ij}}{\sqrt{2 \pi}} \right) 
\exp \left( - \frac{1}{2}A_{ij}^2\right)\equiv  {\rm D}A\ {\rm
  e}^{-\frac12\Tr AA^T}. 
\end{equation}
The eigenvalues $\la_i$ obey the equation $\det [\la_i -A ] = 0$, and
thus are real or occur in complex conjugate pairs. They enjoy the
following ordered joint probability density function (jpdf)
\cite{Lehmann91,Edelman97} 
\begin{equation}
\label{dmuL}
d\mu(\la_1,\la_2, \ldots, \la_N) 
= C_N \cdot  d\la_1 \ldots d\la_N   
\cdot \prod_{i<j}^N (\la_i - \la_j) \cdot \prod_k^N f(\la_k)\ ,   
\end{equation}
with some positive definite weight function $ f(\la_k)=f(\bar\la_k)$ and
a normalisation constant $C_N$. Here the eigenvalues are ordered as follows 
if they are real: $\la_1>\la_2>\ldots$ , if they are complex: 
${\rm  Re}\la_1={\rm Re}\la_2>{\rm Re}\la_3={\rm Re}\la_4>\ldots,
\ {\rm Im}\la_1=-{\rm Im}\la_2>0, 
\ {\rm Im}\la_3=-{\rm  Im}\la_4>0,\ldots$ , 
and similarly if they are mixed (see also \cite{Sommers2008}). 

This implies that the spectral  density of 
complex eigenvalues of the $N+2$ dimensional
ensemble, which can be obtained by inserting a two-dimensional delta-function
in the complex plane, is proportional to 
\begin{equation}
\label{R_2^C}
R_{N+2,\,1}^C(\la) \propto  i(\la-\bar \la) f(\la)^2 
\langle \det[\la-A] \det[\bar\la-A]  \rangle _N \ .
\end{equation} 
The brackets mean the average over the ensemble (\ref{dmuA})
with the partition function 
${\cal Z}_N\equiv \int  {\rm D}A\ \e^{-\frac12\Tr AA^T}$. 
There is an additional contribution to the total spectral density from the
real eigenvalues, which is
obtained by inserting a delta-function on the real axis, 
which we do not consider here. 
We are therefore led to consider the following correlation 
of two characteristic polynomials of $A$ 
\footnote{The Hermitian conjugate  
is put here merely to stress the analogy to other ensembles discussed below, 
as for real matrices $\det[ A^T]=\det [A]$.}
\begin{equation}
\label{F}
 {\cal F }_N (\la, \ga)
\ \equiv\ \langle \det [\la-A] \det[\ga-A^\dag] \rangle _N
\equiv\frac{{\cal K }_N^{\beta=1} (\la ,\ga)}{\la-\ga}\ \
\mbox{with}\ \ \la\neq\ga\ .
\end{equation}
It determines the antisymmetric kernel
${\cal K }_N^{\beta=1} (\la ,\ga)$, 
from which all correlation functions of complex eigenvalues 
follow for even $N$. While the result for odd $N$ is known in the real Ginibre
ensemble \cite{Sommers2008}, 
very recently a general technique has been proposed for obtaining the
odd $N$ result from even $N$ by removing an eigenvalue
\cite{Forrester08II}. 
The kernel in eq. (\ref{F}) has been derived in \cite{Sommers2007} 
using the Edelman result
\cite{Edelman97} for the density of complex eigenvalues, and using
eq.(\ref{R_2^C}). Here we will give an independent derivation which makes
clear why the result is so simple, when the jpdf eq. (\ref{dmuL}) is so
complicated. 

The relation eq. (\ref{F}) is far more general. Not only does it hold for
the other real Ginibre ensembles to be introduced below, but it also holds for
other symmetry classes with complex eigenvalues having 
unitary or symplectic invariance. For the
quaternionic Ginibre ensembles at $\beta=4$ an identical relation to 
eq. (\ref{F}) was shown in \cite{ABa} to give the skew-symmetric kernel.
In the Ginibre ensembles with unitary symmetry $\beta=2$ the kernel is
symmetric and the following modified, simpler relation is known to hold
\cite{AV} 
\be
\label{Fb2}
\langle \det [\la-A] \det[\ga-A^\dag] \rangle _N
={\cal K }_N^{\beta=2} (\la ,\ga)\ .
\ee

The argument we just presented above for the real Ginibre ensemble
at maximal asymmetry can easily be translated to the partially symmetric 
case depending on an asymmetry parameter $\tau$. 
Here in the large-$N$ limit
the complex eigenvalues lie inside an ellipse with axes $\sim(1\pm\tau)$
\cite{Sommers88}. 
It is known \cite{Lehmann91} that 
its jpdf is related to eq. (\ref{dmuL}) by a simple
rescaling of the eigenvalues, and we readily obtain
\begin{equation}
\label{Ft}
\fl {\cal F }_N (\la, \ga;\tau)
\ \equiv\ \langle \det [\la-(S+vA)] \det[\ga-(S+vA)^T] \rangle _N
\equiv\frac{{\cal K }_N^{\beta=1} (\la ,\ga;\tau)}{\la-\ga}\ ,
\end{equation}
with 
\be 
\label{tvdef}
\tau\in[0,1]\ ,\ \ v^2=\frac{1-\tau}{1+\tau}\ .
\ee
The average is with respect to the following partition function
\be
\label{Zt}
{\cal Z}_N\ \equiv\ \int {\rm D}S\  {\rm D}A\ 
\exp\left[-\frac{1}{2(1+\tau)}\Tr  (SS^T+AA^T)\right]\ ,
\ee
and we consider the eigenvalues of the partly symmetric matrix $J=S+vA$. Here 
$S$ and $A$ are $N\times N$ matrices being symmetric and antisymmetric
respectively, with a particular choice of variance. 
The limiting case $\tau=0$ brings us back to the ensemble eq. (\ref{dmuA})
while setting $\tau=1$ would lead to the Gaussian Orthogonal Ensemble. 
However, in that case the eigenvalues become real and this limit is
subtle.

The second ensemble we consider in this paper is the chiral counterpart 
of the real Ginibre ensemble. Following \cite{HOV} and its extension to a
two-matrix model \cite{James} we define the following chiral real Ginibre
ensemble ($ch$) with a particular variance $n$, 
\be
\label{Zchmu}
{\cal Z}_N^{ch}\ \equiv\ \int {\rm D}A\  {\rm D}B\ 
\exp\left[-\frac{n}{2}\Tr (AA^T+BB^T)\right]\ .
\ee
Again we compute the average of characteristic polynomials to obtain the
kernel, 
\bea
\label{Kchdef}
&&{\cal F }_N^{ch} (\la, \ga;\mu)
\equiv \langle 
\det[\la-M]\det[\ga-M^T]
\rangle _N\ \equiv\ 
\frac{{\cal K }_N^{ch,\,\beta=1} (\la ,\ga;\mu)}{\la^2-\ga^2}\ ,
\\
&&M\equiv \left(
\begin{array}{cc}
0& A+\mu B\\
A^T-\mu B^T & 0\\
\end{array}
\right) .
\label{Mdef}
\eea
Here both $A$ and $B$ are rectangular $N\times (N+\nu )$ 
matrices without further symmetry among the real matrix elements. They are 
drawn independently from the ensemble (\ref{dmuA}) extended to $\nu\geq0$.
The asymmetry parameter is given here by $\mu\in[0,1]$, where $\mu=1$ denotes
maximal asymmetry, and $\mu=0$ takes us back to the chiral Gaussian Orthogonal
Ensemble. In \cite{HOV} initially a one-matrix model was proposed, replacing
$B$ by the identity. Whilst we expect that in the large-$N$ limit both lead to
the same universal result our choice allows for an eigenvalues basis, as in
the corresponding chiral extensions of Ginibre at 
$\beta=2$ \cite{James} and $\beta=4$
\cite{A05}, having complex and quaternion real matrix elements, respectively.
The ensemble in eq. (\ref{Zchmu})
has not been solved before and we will give a  completely new result
below, depending parametrically on $\nu$. 
For maximal asymmetry it corresponds to class $2P$ in \cite{Magnea08} for real
elements.

The $2N+\nu$ eigenvalues $\la_i$ of the matrix $M$ defined in eq. (\ref{Mdef})
satisfy
\be
\label{chev}
0\ =\ \det[\la-M]\ =\ \la^\nu\det\left[\la^2-(A+\mu B)(A^T-\mu B^T)
\right]\ .
\ee
It will be shown elsewhere that the jpdf of these eigenvalues
is again of the form in eq. 
(\ref{dmuL}).  From that it follows that the kernel 
derived from ${\cal F}_N^{ch} (\la, \ga;\mu)$ again determines all correlation 
functions of complex eigenvalues. 

A peculiarity of the chiral ensemble is the following: the non-zero
eigenvalues $\la^2_i$ solving the second equation in (\ref{chev}) are real but
not necessarily positive. Thus the eigenvalues of $M$ can have both real and
purely imaginary eigenvalues as well as complex conjugate
eigenvalue pairs. 
Moreover, all non-zero eigenvalues come in $\pm$ pairs due to the
chirality of the matrix $M$.

\section{Characteristic polynomials for the real Ginibre ensemble}
\label{Gin1}

For pedagogical reasons we begin with the maximally asymmetric case
eq. (\ref{F}). The partly symmetric case at $\tau\neq 0$ is given as a second
example below.
\be
\label{F_1}
 {\cal F }_N(\la, \ga)= \frac{1}{{\cal Z}_N}
\int {\rm D}A\ {\rm e}^{-\frac12{\rm Tr} A A^T}
 \det[\la-A]\ \det[\ga - A^T] \ . 
\ee
Writing the determinants in terms of two $N$-dimensional complex 
Grassmann vectors  $\eta_i$ and $\zeta_i$, with $i=1,\ldots,N$, we obtain 
\bea
\label{F_A2}
\fl 
{\cal F }_N (\la, \ga)&=& \frac{1}{{\cal Z}_N}
\int {\rm D}A \int d\zeta\ d\eta\  
\exp\left[-\frac12A_{ij}A^T_{ji} -\la\zeta^*_i\zeta_i - \ga
\eta^*_i\eta_i+\zeta^*_iA_{ij}\zeta_j +\eta^*_jA^T_{ji}\eta_i\right] \nn\\ 
\fl &=&   \int d\zeta\ d\eta\  
\exp\left[     -\la\zeta^*_i\zeta_i - \ga \eta^*_i\eta_i+
\frac12(\zeta_i^*\zeta_j+\eta_j^*\eta_i)^2\right] \ ,
\eea
after integrating out the Gaussian matrix $A$. 
Here and in the following we will use summation conventions over double
indices. 
The last term in the exponent can be written as
\begin{equation}
\label{F_A4}
\frac12 (\zeta_i^*\zeta_j+\eta_j^*\eta_i)(\zeta_i^*\zeta_j+\eta_j^*\eta_i)
=\zeta_i^*\eta_i\ \zeta_j\eta_j^* \ .
\end{equation}
With the help of a complex Hubbard-Stratonovich (HS) transformation 
we can bilinearise and integrate out the Grassmann variables:
\bea
\label{F_A5}
\fl 
{\cal F }_N (\la,\ga)&=&\frac{1}{\pi}\int{ d^2z}  
\int d\zeta\ d\eta\ 
\exp\left[-|z|^2    -\la\zeta^*_i\zeta_i - \ga \eta^*_i\eta_i
+z\zeta_i\eta_i^* +\bar z \zeta_i^*\eta_i \right] \nn\\ 
&=& \frac1\pi\int d^2z\ {\rm e}^{-|z|^2}\ (\la\ga+|z|^2)^N\ =\ 
N!\sum_{n=0}^N{(\la\ga)^{n}\over n!}    \ . 
\eea
This gives a polynomial with leading power $(\la\ga)^N$ as expected. 
Thus our first main result leads to the following antisymmetric kernel 
\be
\label{kernel1}
{\cal K }_N^{1} (\la ,\ga)\ =\ (\la-\ga)\ 
N!\sum_{n=0}^N{(\la\ga)^{n}\over n!} \ ,
\ee
which is enough to derive all complex correlation functions. 
On setting $\ga=\bar\la$ and multiplying by the weight $f(\la)^2$, 
Edelman's complex density \cite{EKS94} in terms of an incomplete exponential 
follows. It is
remarkable that it only depends on $|\la|^2$ while the jpdf eq. 
(\ref{dmuL}) is not isotropic. We note that Edelman derived his result using
methods from multivariate statistics, and not from the jpdf.

We now turn to the partly symmetric case with $\tau\in[0,1]$, where we can
follow the same path, 
\bea
\fl{\cal F }_N (\la, \ga;\tau)
&\!\!\!=&\!\frac{1}{{\cal Z}_N}\int\! {\rm D}S\,  {\rm D}A\ 
\e^{-\frac{1}{2(1+\tau)}\Tr  (SS^T+AA^T)}
\det [\la-(S+vA)] \det[\ga-(S+vA)^T] \nn\\
&\!\!\!=&\!\frac{1}{{\cal Z}_N}
\int\! {\rm D}S\, {\rm D}A\int d\zeta\ d\eta\ 
\exp\Big[-\frac{1}{2(1+\tau)}(S_{ij}^2-A_{ij}^2) 
-\la\zeta^*_i\zeta_i - \ga\eta^*_i\eta_i\nn\\
&&\ \ \ \ \ \ \ \ \ \ \ \ \ \ \ \ \ \ \ \ \ \ \ \ \ \ \ \ \ \ \ \ \ \ \ 
+\zeta^*_i(S_{ij}+vA_{ij})\zeta_j +\eta^*_i(S_{ij}-vA_{ij})\eta_j\Big].
\eea
After symmetrising and antisymmetrising the terms in the last line, e.g. 
$\zeta^*_iS_{ij}\zeta_j=\frac12S_{ij}(\zeta^*_i\zeta_j+\zeta^*_j\zeta_i)$, we
can complete the squares in $S_{ij}$ and $A_{ij}$ respectively, and integrate
them out to obtain
\bea
\label{FtnoAS}
\fl{\cal F }_N (\la, \ga;\tau)
&=&\int d\zeta\ d\eta\ \exp\Big[-\la\zeta^*_i\zeta_i - \ga\eta^*_i\eta_i
-c_-^2((\zeta^*_i\zeta_i)(\zeta^*_j\zeta_j)
+(\eta^*_i\eta_i)(\eta^*_j\eta_j)) \nn\\
&&\ \ \ \ \ \ \ \ \ \ \ \ \ \ \ \ \ \ \ \ \ \ \ \ \ \ \ \ \ \ \ \ \ \ 
\ \ \ \ \ \ \ \ \ \ \ 
-2c_+^2\eta^*_i\zeta^*_i\eta_j\zeta_j +2c_-^2\eta^*_i\zeta_i\eta_j\zeta_j^* 
\Big]\ .
\eea
Here we have introduced the constants $c_\pm^2\equiv\frac12(1+\tau)(1\pm
v^2)$.
The quartic terms in the Grassmann variables can be rewritten using two real 
HS transformations for the first line of eq. (\ref{FtnoAS}), and two complex
ones for the second line:
\bea
&&\fl{\cal F }_N (\la, \ga;\tau)
=\frac{1}{\pi^3}\int dx\ dy \int d^2z\ d^2w
\int d\zeta\ d\eta\ \exp\Big[-x^2- y^2 
-\zeta^*_j(\la+2ic_-x)\zeta_j 
\nn\\
&&\fl\ \ \ \ \ \ \ \ \ \ \ -\eta^*_j(\ga+2ic_-y)\eta_j -2|z|^2-2|w|^2
+2c_+(\bar z \eta_j\zeta_j-z\eta_j^*\zeta_j)
-2c_-(\bar w \eta_j\zeta_j^*+w\eta_j^*\zeta_j)
\Big]\nn\\
&&\fl\ \ \ \ \ \ \ \ \ \ \ \ \ \ =\frac{1}{\pi^3}\int dx\ dy \int d^2z\ d^2w
\ \exp\Big[-x^2- y^2 -2|z|^2-2|w|^2\Big]\nn\\
&&\fl\ \ \ \ \ \ \ \ \ \ \ \ \ \ \ \ \ \ \times\Big[ 
(\la+2ic_-x)(\ga+2ic_-y)+4c_+^2|z|^2+4c_-^2|w|^2\Big]^N\ .
\label{FtHS}
\eea
Expanding the last factor twice into binomial series in powers of $|z|^2$ and 
$|w|^2$ we can apply the following integral representation of the Hermite
polynomials: 
\be
\label{Hint}
\left( \frac{\tau}{2}\right)^{\frac{k}{2}}
H_k\left( \frac{\la}{\sqrt{2\tau}}\right)=\frac{1}{\sqrt{\pi}}
\int dx \ \e^{-x^2} \left(\la+i\sqrt{2\tau}x\right)^k\ ,
\ee
where using eq. (\ref{tvdef}) we have $2c_-=\sqrt{2\tau}$. 
This eliminates the two real integrations.
After integrating out the two remaining complex variables $z$ and $w$ we
finally arrive at
\be
\label{Ftfinal}
{\cal F }_N (\la, \ga;\tau)=N!\sum_{l=0}^N\tau^l\sum_{k=0}^l
\frac{1}{k!\ 2^k}H_k\left( \frac{\la}{\sqrt{2\tau}}\right)
H_k\left( \frac{\ga}{\sqrt{2\tau}}\right)\ .
\ee
As a check this is again a polynomial with leading order $(\la\ga)^N$. 
Although our result eq. (\ref{Ftfinal}) could be further simplified 
this form is most useful for obtaining the antisymmetric kernel by applying the
Christoffel-Darboux formula to the inner sum:
\be
\fl {\cal K }_N^{1} (\la ,\ga;\tau) \ = \ N!\sum_{l=0}^N\frac{1}{l!}
\left( \frac{\tau}{2}\right)^{l+\frac{1}{2}}\left( 
H_{l+1}\left( \frac{\ga}{\sqrt{2\tau}}\right)
H_l\left( \frac{\la}{\sqrt{2\tau}}\right)
-(\ga\leftrightarrow\la)\right)
\ .
\ee
This coincides precisely with the kernel of skew-orthogonal Hermite
polynomials derived in \cite{Forrester08} via the jpdf, which is much more
elaborate. As was shown there independently, this kernel generates all
complex eigenvalue correlation functions of the partly symmetric ensemble eq. 
(\ref{Zt}) for even $N$, depending parametrically on $\tau$.

A similar kernel given in terms of orthogonal (for $\beta=2$) \cite{FKS98} 
or skew-orthogonal (for $\beta=4$) \cite{EK01}
Hermite polynomials is known for the partly symmetric Ginibre ensembles.

\section{Characteristic polynomials for the chiral real Ginibre ensemble}
\label{chGin1}

In this section we present the calculation only for the 
partly symmetric case of the chiral extension of the real Ginibre 
ensemble, depending on asymmetry parameter $\mu$.
The simpler result at
maximal asymmetry with $\mu=1$ is given at the end of this section. 
\bea
\fl{\cal F }_N^{ch} (\la, \ga;\mu)
&=&\frac{1}{{\cal Z}_N^{ch}}\int \!\!{\rm D}A\,  {\rm D}B\ 
\e^{-\frac{n}{2}\Tr  (AA^T+BB^T)}
\det \left[
\begin{array}{cc}
\la& -(A+\mu B)\\
-(A^T-\mu B^T) & \la\\
\end{array}
\right]\nn\\ 
&&\ \ \ \ \ \ \ \ \ \ \ \ \ \ \ \ \ \ \ \ \ \ \ \ \ \ \ \ \ \ \ \ \ \ \ \,
\times \det\left[
\begin{array}{cc}
\ga          & -(A-\mu B)\\
-(A^T+\mu B^T) & \ga\\
\end{array}
\right] \nn\\
&&\fl\!\!\!\!\!\!\!\!\!=\frac{1}{{\cal Z}_N^{ch}}
\int \!\!{\rm D}A\,  {\rm D}B\int \!\!d\eta d\psi d\zeta d\vp\ 
\exp\Big[-\frac{n}{2}(A_{ia}^2+B_{ia}^2) 
-\la(\eta^*_i\eta_i +\psi_a^*\psi_a)- \ga(\zeta^*_i\zeta_i+\vp_a^*\vp_a) \nn\\
&&\fl\!\!
+\eta^*_i(A_{ia}+\mu B_{ia})\psi_a +\psi_a^*(A_{ai}^T-\mu B_{ai}^T)\eta_i
+\zeta_i^*(A_{ia}-\mu B_{ia})\vp_a +\vp_a^*(A_{ai}^T+\mu B_{ai}^T)\zeta_i
\Big].\ \ 
\eea
Here we have written each determinant of size $2N+\nu$ in terms of {\it two}
Grassmann vectors, $\eta_i$($\zeta_i$) and $\psi_a$($\vp_a$) 
of size $N$ and $N+\nu$,
respectively. Our summation conventions imply for $i=1,\ldots,N$, and 
for $a=1,\ldots,N+\nu$. 
In addition to $\mu$ we have a parameter $n$ for the
variance. After completing the square and integrating out the matrices $A$ and
$B$ we obtain
\bea
\fl{\cal F }_N^{ch} (\la, \ga;\mu)&=& \int d\eta d\psi d\zeta d\vp\ 
\exp\Big[-\la(\eta^*_i\eta_i +\psi_a^*\psi_a)
-\ga(\zeta^*_i\zeta_i+\vp_a^*\vp_a) \nn\\
&&
\fl+\frac{1}{2n}(\eta_i^*\psi_a-\eta_i\psi_a^*+\zeta_i^*\vp_a-\zeta_i\vp_a^*)^2
+\frac{\mu^2}{2n}
(\eta_i^*\psi_a+\eta_i\psi_a^*-\zeta_i^*\vp_a-\zeta_i\vp_a^*)^2
\Big]\ .
\label{FchGrass}
\eea
Multiplying out and collecting all nonzero terms we need 6 new complex 
integrations to perform the HS transformations that bilinearise the Grassmann
variables. We only give the result obtained after performing all Grassmann
integrations:
\bea
\label{FHS}
\fl{\cal F }_N^{ch} (\la, \ga;\mu)&=& \frac{1}{\pi^6}
\int d^2u\ d^2v\ d^2w\ d^2p\ d^2q\ d^2z
\ \e^{-|u|^2 -|v|^2-|w|^2-|p|^2-|q|^2-|z|^2}\\
&&\fl\!\!\!\!\!\!
\times \left(
(\la-i\delta_-u)(\ga-i\delta_-\bar v)
-\delta_+^2wz+\delta_-^2pq\right)^N 
\left( 
(\la-i\delta_-\bar u)(\ga-i\delta_-v)
-\delta_+^2\bar w\bar z
+\delta_-^2\bar p\bar q \right)^{N+\nu}\!\!,\nn
\eea
where we have used the following abbreviations
\be
\delta_\pm^2 \equiv \frac1n (1\pm\mu^2)\ .
\label{deltadef}
\ee
Expanding the first factor in the second line of eq. (\ref{FHS}) as 
\be
\fl(\hat{\la}\hat{\bar\ga}\ -\ \delta_+^2wz\ +\ \delta_-^2pq)^N=\sum_{l=0}^N
{N\choose l}( -\delta_+^2wz)^{N-l}\sum_{k=0}^l{l \choose k} 
(\hat{\la}\hat{\bar\ga})^k (\delta_-^2pq)^{l-k}\ , 
\ee
with $\hat{\la}\hat{\bar\ga}=(\la-i\delta_-u)(\ga-i\delta_-\bar v)$,
and likewise the second factor, we can use the following orthogonality
relation
\be
\frac1\pi \int d^2p\ \e^{-|p|^2} p^k {\bar p}^{\,l} \ = \ \delta_{kl}k! \ .
\ee
Applying this first to the integrations over variables 
$p$ and $q$, and then to $w$
and $z$ we can reduce the four sums to two. As a final step we employ the
following complex integral representation for Laguerre polynomials
\be
\frac1\pi \int d^2u\ \e^{-|u|^2} (\la+iu)^k(\la+i\bar u)^{k+\nu}
\ =\ k!\ (-)^k\la^\nu
L_k^\nu(\la^2) \ . 
\label{Lknurep}
\ee
Whilst we did not find this representation in tables it can be easily verified
from the standard representation of generalised Laguerre polynomials
\be
L_k^\nu(x)\ =\ \sum_{m=0}^k \frac{(-1)^m(k+\nu)!}{(k-m)!\,(m+\nu)!\,m!}\ x^m\ .
\ee
Using the integral representation eq. (\ref{Lknurep}) as well as its complex
conjugate we finally arrive at the following result:
\be
\fl{\cal F }_N^{ch} (\la, \ga;\mu)\ =\ N!\,(N+\nu)!\,\delta_+^{4N}(\la\ga)^\nu
\sum_{l=0}^N\left( \frac{\delta_-}{\delta_+}\right)^{4l}
\sum_{k=0}^l \frac{k!}{(k+\nu)!} L_k^\nu\left( \frac{\la^2}{\delta_-^2}\right)
L_k^\nu\left( \frac{\ga^2}{\delta_-^2}\right).
\label{Fchfinal}
\ee
It is a polynomial in $\la$ and $\ga$ with the correct leading power 
$(\la\ga)^{2N+\nu}$. 
Looking back to the definition of the antisymmetric 
kernel in our chiral case, eq. 
(\ref{Kchdef}), we can read off the following result, after 
using the Christoffel-Darboux formula for Laguerre polynomials:
\bea
\fl{\cal K }_N^{ch,\,1} (\la ,\ga;\mu)&=&
N!\,(N+\nu)!\ \delta_+^{4N}(\la\ga)^\nu\nn\\
\fl&&\times
\sum_{l=0}^N\left( \frac{\delta_-}{\delta_+}\right)^{4l}
\frac{(l+1)!}{(l+\nu)!}\ \delta_-^2
\left(\! 
L_{l+1}^\nu\Big( \frac{\ga^2}{\delta_-^2}\Big)
L_{l}^\nu\Big( \frac{\la^2}{\delta_-^2}\Big)
-(\ga\leftrightarrow\la)\!
\right)\!.
\label{Kchfinal}
\eea
This gives our new kernel of the chiral real Ginibre ensemble, from which all
its complex eigenvalue correlations follow. In particular for 
$\ga=\bar\la$ it is proportional to a new complex eigenvalue
density as in eq. (\ref{R_2^C}).
It is similar to the corresponding 
expressions for the kernel at $\beta=2$ \cite{James} 
and $\beta=4$ \cite{A05}, also given 
in terms of Laguerre polynomials in the complex
plane. 
 
After dealing with the general case we can go to maximal asymmetry, by
setting $\mu=1$. In this limit only the leading power of the Laguerre
polynomials contributes, and we obtain 
\be
\fl{\cal F }_N^{ch} (\la, \ga;\mu=1)\ =\ N!\,(N+\nu)!
\left( \frac2n\right)^{2N}(\la\ga)^\nu
\sum_{k=0}^N\frac{1}{k!\,(k+\nu)!}\left(\frac{n\la\ga}{2}\right)^{2k}
\label{Fchmax}
\ee
for the characteristic polynomials, with $\lim_{\mu\to1}\delta_+^2=\frac2n$. 
For the corresponding kernel we have to properly rescale with $\delta_-^2$ 
and we obtain
\be
\fl\lim_{\mu\to1}\delta_-^2{\cal K }_N^{ch,\,1} (\la ,\ga;\mu)=
N!\,(N+\nu)!
\left( \frac2n\right)^{2N}(\la\ga)^\nu (\la^2-\ga^2)
\sum_{l=0}^N\frac{1}{l!\,(l+\nu)!}
\left(\frac{n\la\ga}{2}\right)^{2l}\!.
\label{Kchmax}
\ee
When setting $\ga=\bar\la$ and comparing to eq. (\ref{kernel1}) 
we again find a dependence on the modulus only, despite the anisotropic jpdf.
Here the incomplete exponential is replaced by an incomplete modified
$I$-Bessel function of the first kind.

\section{Conclusions}
\label{concs}

We have calculated the expectation value of the product 
of two characteristic polynomials 
with respect the following two Gaussian random matrix models: 
the partly symmetric real Ginibre ensemble, and its chiral counterpart, 
a newly introduced two-matrix model. 
In our calculation we have used the supersymmetric method, 
without the need to explicitly go to an eigenvalue basis. 
In this simple way we can determine a skew-symmetric kernel which is  
the main building block for all 
complex eigenvalue correlation functions that can be written as
Pfaffians. One could calculate this kernel directly from
the joint eigenvalue distribution (jpdf), 
but this turns out to be a very difficult task.

This kernel
is given by a sum over Hermite polynomials for the real Ginibre case, 
depending on the asymmetry parameter. Here we have recovered a known, very 
recent result. In the chiral real Ginibre ensemble we find a new kernel given
in terms of generalised 
Laguerre polynomials. In addition to the asymmetry $\mu$ 
it depends on the parameter
$\nu$ labelling the number of exact zero eigenvalues.
Our method offers an explanation of why the spectral density of complex
eigenvalues is so simple, i.e. being an incomplete exponential or $I$-Bessel
function at maximal asymmetry, while the jpdf is so complicated. 

One possible application of our new chiral result would be in field theory 
for Dirac operators with a real representation. The reason complex
eigenvalues appear here is due to a chemical potential $\mu$ of the quarks. 

It is an open question for the chiral ensemble if for all $N$ the kernel 
also determines the weight function $f(\lambda)$, and if both
ingredients (i.e. kernel and weight) determine all correlation functions 
of real, complex and mixed eigenvalues. For the real Ginibre ensemble 
this fact is known to hold, and the similarity in structure makes this very
suggestive. 
\\

H.-J. S. acknowledges discussions with  D. Savin as well as the kind
hospitality at Brunel University   with thanks. This work has been supported by
EPSRC grant EP/D031613/1, European Network ENRAGE MRTN-CT-2004-005616 (G.A.), 
an EPSRC doctoral training grant (M.J.P.) and the 
SFB/TR12 of the Deutsche Forschungsgemeinschaft (H.-J.S.).

\end{document}